\documentstyle[aps,multicol,graphicx]{revtex}

\begin{document}

\def\ve{\varepsilon}

\title{Electrons on a sphere in disorder potential} 
\author{D.N. Aristov $^{1,2}$}
\address{ $^1$ Petersburg Nuclear Physics Institute,
   Gatchina, St. Petersburg 188350, Russia \\
$^2$ NORDITA, Blegdamsvej 17, DK-2100, Copenhagen, Denmark}

\date{Dec 14, 2000} 
\maketitle

\begin{abstract} 
We investigate, both analytically and numerically, the behavior of the
electron gas on a sphere in the presence of point-like impurities. 
We find a criterion when the disorder can be regarded as small one and the main
effect is the broadening of rotational multiplets. In the latter regime the
statistics of one impurity-induced band is studied numerically.  The energy
level spacing distribution function follows the law $P(s) \sim s \exp(-a s^b)$
with $1<b<2$. The number variance shows various possibilities, strongly
dependent on the chosen model of disorder. 
\end{abstract}

 \pacs{ 
73.20.-r, 
72.10.Fk, 
05.45.Pq, 
73.20.Fz 
}

\begin{multicols}{2} \narrowtext

\section{introduction} 

A considerable amount of the theoretical interest has been devoted last years
to the investigation of electronic properties of the nano-size objects
possessing regular cylindrical and spherical shapes. While most of these
studies were consecrated to the various properties of carbon nanotubes,  
the technological advances in production of spherical objects also
require  further theoretical understanding of peculiar quantum effects of
electronic motion in such  materials. 

A well-known fullerene molecule, C$_{60}$, could be one of the  examples,
along with its further modification, so-called ``onion" graphitic structures.
\cite{onion}  The spherical nano-size objects are also found in the studies
of the nonlinear optical response in composite materials \cite{composites} and 
of simple metal clusters. \cite{Ekardt}  
A rapidly evolving field of photonic-band-gap materials \cite{PBG} provides yet
another example  of spheres of about  300 nm in diameter. In the latter case,
one may find silica balls with semiconducting coating (coated opals)
\cite{opals}, or ``inverse'' opal structures, where the initial SiO$_2$
template is  chemically removed  and carbon spherical shells form
three-dimensional fcc structure. \cite{zakhidov}

In all these cases one can assume that the motion of itinerant electrons is
confined within a spherical layer of a width small comparing to a radius. 
To a first approximation, one ignores the interaction effects and  considers a
situation of the electron gas.  Theoretical efforts in this direction comprise
an analysis of the behavior of such gas on a sphere in the  uniform magnetic
field and an evaluation of the electronic correlations without a field. 
It was shown particularly that if the field is small so that the magnetic
length is larger than the radius,   then one can expect the jumps in the
magnetization and susceptibility of the sphere.  \cite{sphere-field,sphereM}
An extension of this analysis for the  ellipsoid of revolution
has been undertaken recently. \cite{bulaev,malits}
In stronger fields, when the magnetic length is smaller than radius, one
finds a series of interesting effects described elsewhere. \cite{FLP,sphereM}

The electronic correlation functions for the topology of the sphere  exhibit,
particularly, a peculiar coherence effect when the coherence length of
electronic motion exceeds the radius of the sphere.  \cite{sphereC} It turns
out, that the amplitude of these functions is enhanced for the antipodal
points on the sphere, where all partial waves of quantum motion come in phase. 

The electron gas approximation, employed in these works, should be violated in
experimental realizations of nanospheres.  One of the reasons for this
violation is  the effects of electron-electron interaction, which were studied
in \cite{ee-inter}.  

Another source of possible inadequateness of the electron gas model is the
explicit absence of rotational symmetry, inferred by the presence of
impurities.  Indeed, the experimental realization of the spherical layer,
wherein the electrons are confined, might be far from  the ideal shape. The
inhomogeneities of various kind (impurities, fluctuations of layer's width)
should eventually destroy  the effects, found theoretically  for the idealized
model.  As a first crucial step here, one observes that the impurities 
necessarily raise the multiple degeneracy of the energy levels, found initially
in a quantum rotator model of a  free gas. 

In this paper we consider the effects of potential impurity  scattering  in
the electron gas on a sphere. We employ the model of point-like impurities,
which  is shown to be valid if the angular momentum of an electron
does not exceed inverse angular range of the potential.  Our results suggest
that there is a definite range of parameters, where the electron gas
approximation should be applicable. Particularly,  this
approximation is validated, when the radius of the sphere  is smaller
than the ratio $\sqrt{\nu} /\nu_{imp}$, with 
the areal densities of electrons and impurities, $\nu$ and $\nu_{imp}$,
respectively.  We show that in this case the degeneracy of the initially
degenerate multiplet, close to Fermi energy, is lifted only partially. The
unsplit states are  superpositions  of the initial spherical harmonics, and
hence describe the electrons freely propagating along the sphere. As a result,
one expects, that for sufficiently small disorder it is possible to discuss
the coherence effects, induced by the spherical topology.   

For the regime of weak disorder, we numerically investigated the splitting of 
on multiplet, caused by random impurities. 
We found that the statistics of
the energy levels in this case is not identical, albeit close, to the
predictions of the random matrix theory for the Gaussian orthogonal ensemble. 
Particularly we show that the energy level spacing distribution follows the
law $P(s) \sim s \exp - s^b$, with $b\simeq 1.76$ in the limit of white-noise
potential distribution.  We discuss a possible reason for this deviation,
related to the geometry of the problem. 

A rest of the paper is organized as follows. 
We formulate the problem and treat the random potential perturbatively in
Sec.\  II. We identify here the region of parameters, where the perturbation
theory fails.  In Sec.\  III we numerically investigate this region. The
discussion and conclusions are presented in Sec.\   IV.

\section{perturbative treatment of the impurity potential}

\subsection{Green's function}

In the ideal situation the motion of an electron on the sphere is described by
the quantum rotator model. 
 The wave functions $\Psi $ 
are the spherical harmonics $Y_{lm}$ and the
spectrum is simple  :
        \begin{equation}
        \Psi(\theta,\phi) = r_0^{-1} Y_{lm}(\theta,\phi), \quad
        E_{l} = (2m_er_0^2)^{-1} l(l+1).
        \label{spectrum}
        \end{equation}
We normalize $\Psi(\theta,\phi)$ on the surface of the sphere
$r_0^2 \int |\Psi|^2\sin\theta d\theta d\phi  =1 $.

The main effect of the random impurity potential is the lifting of
$(2l+1)-$fold degeneracy of each $l-$th energy level.  Qualtitatively, one 
expects  that at sufficiently small disorder  these split levels form a 
subband nearly the initial position $E_l$, and the width of this subband does
not exceed the separation between the adjacent levels $E_l$ and $E_{l\pm1}$. 
This regime  corresponds to the mean free path of the electron, $l_{mfp}$, 
larger than $r_0$. The coherence effects of the electronic motion should be
still present in this case. \cite{sphereM,sphereC} 

At larger disorder, the width of induced subband is more than $|E_l
- E_{l\pm1}|$. The different subbands overlap now and, as a result, one
finds a constant density of states, typical for a planar two-dimensional
electron gas.   Having the relation $l_{mfp} < r_0$ in this case, one sees
that the spherical surface is effectively decomposed onto the ``patches'' of
size $l_{mfp}$, in which the electronic motion is essentially planar. 

To find out the corresponding criterion for these two regimes, we use 
the Green function formalism.  Treating disorder as perturbation, we write the
self-consistent equation for the Green function. The spread of the split
levels around the position of the initial multiplet is described by an
imaginary part of the corresponding self-energy. This approach is
very similar to one employed in the analysis of splitting of the lowest Landau
level in two-dimensional electron gas (2DEG), \cite{Ando,baskin} and a reader
can easily find parallels with our situation below.

The Green function is given by 
     \begin{equation}
     G(\omega) =
     \sum_{lm}\frac{ \Psi_{lm}^\ast(\theta',\phi) \Psi_{lm}(\theta,0)
     }{\omega +\mu -E_{lm}} ,
     \label{defG}
     \end{equation}
For the free motion this expression simplifies
     \begin{equation}
     G^0(\omega) =
     (4\pi r_0^2)^{-1}\sum_{l}\frac{ (2l+1) P_l(\cos\Omega)
     }{\omega +\mu -E_{l}} ,
     \label{G0series}
     \end{equation}
with the distance
between two points on the sphere 
     $
     \cos\Omega =
     \cos\theta \cos\theta' +
     \sin\theta \sin\theta' \cos\phi.
     $

The basic equation is
     \begin{equation}
     G(\Omega_{12}) =
     G^0(\Omega_{12}) +
     \int d{\bf r}_3 d{\bf r}_4
     G^0(\Omega_{13})
     \Sigma(\Omega_{34})
     G(\Omega_{42})
     \end{equation}
where we write $\Omega_{12}$ for the angular distance between
$\Omega_{1}$ and $\Omega_{2}$, etc.
The integration over the surface of the sphere reads as
$\int d{\bf r} = r_0^2 \int d\Omega$.

Being averaged over disorder, the
functions entering the above equation depend only on the distances
between the corresponding points. Representing each function
through its generalized Fourier coefficients, 
         \begin{equation}
         F(\Omega) = r_0^{-2} \sum_{l} \frac{2l+1}{4\pi} F_{l} P_{l}(\Omega),
         \label{Fourier}
         \end{equation}
we have
     \begin{equation}
     G_l^{-1} =
     (G_l^0)^{-1} -
     \Sigma_l
     \end{equation}
with $G_l^0 = (\omega - E_l)^{-1}$.  We consider different contributions to
$\Sigma_l$ below. But before doing that, we list some formulas applicable to
the spherical geometry in the next subsection. 

\subsection{some useful formulas}

First we note that eq. (\ref{Fourier}) allows other representations, namely
        \begin{eqnarray}
         F(\Omega_{12}) &=& r_0^{-2} \sum_{l} F_{l} A_l Y_{l0}(\Omega_{12}),
         \nonumber \\ &=& 
         r_0^{-2} \sum_{lm} F_{l} Y_{lm}^\ast(\Omega_1)Y_{lm}(\Omega_2),
         \end{eqnarray}
with the quantity 
        \begin{equation}
        A_l \equiv \sqrt{(2l+1)/(4\pi)}.
        \end{equation}

\noindent
The Green function (\ref{G0series}) can be represented
\cite{sphereM} through the Legendre function
     \begin{equation}
     G^{0}(\omega) =
     -\frac{m_e}{2\cos \pi \lambda }
     P_{-1/2+\lambda }(-\cos\Omega),
     \label{G0}
     \end{equation}
with  $\lambda = \sqrt{2m_er_0^2(\mu+\omega)+1/4}$.   In subsequent
consideration we are mostly interested in energies close to Fermi level,
$\omega \ll \mu$.  In this sense, we define the angular Fermi momentum, $L$, 
as $E_L \simeq \mu$, more precisely 
     \begin{eqnarray}
     \lambda  &\simeq&
     L+1/2+(\omega-\Sigma_l)/\Delta E
     \label{lambda}
     \\
     \Delta E &=& E_{L+1}-E_L = L/(m_er_0^2).
     \end{eqnarray}

For small angular distances $\Omega \ll \lambda^{-1}$ we have
     \begin{equation}
     G^{0}(\omega) \simeq
     \frac{m_e}{2\pi}
     \left[
     2 \ln\left(
     \lambda  \sin \frac\Omega2\right) +\gamma -
     \pi \tan\pi \lambda
     \right]
     \label{G0small}
     \end{equation}
with the Euler constant
$\gamma = 0.577\ldots$.
This expression is similar to the Green function
for the 2DEG in the magnetic field $B$. Indeed, in the latter case we
have for $r \ll l_\ast$ \cite{GeyMar}

     \begin{equation}
     G^{0}(\omega) \simeq
     \frac{m_e}{2\pi}
     \left[
     2 \ln\left(
     \frac{r}{l_\ast} \sqrt{\frac{\omega}{2\omega_c}}
     \right) -2\gamma -
     \pi \tan\pi
     \frac{\omega}{\omega_c}
     \right]
     \label{G0LL}
     \end{equation}
where the magnetic length $l_\ast = (eB)^{-1/2}$ and the
cyclotron frequency $\omega_c=eB/m_e$.

For larger distances, $\lambda \sin \Omega \agt 1$  
we have approximately 
     \begin{equation}
     G \simeq
     - \frac{m_e}{\sqrt{2\pi \lambda \sin\Omega}}
     \frac{\cos[\lambda(\pi-\Omega)-\pi/4]}{\cos \pi \lambda}  ,
     \label{G0-asymp}
     \end{equation}
As was shown in \cite{sphereM}, the existence of two oscillating exponents in
(\ref{G0-asymp}) ,  $\exp\pm i\lambda \Omega_{12}$,  indicates the
interference of the partial waves on the sphere, one wave going along the
shortest way between the the points $\Omega_1$ and $\Omega_2$, and another
going along the longest way, turning around the sphere.

We discuss now a notion of point-like impurity on the sphere. 
Letting first the range of the potential to be zero, we write 
$u_0 \delta({\bf r}-{\bf
r}_i)= u_0 r_0^{-2} \delta(\Omega- \Omega_i) \equiv u_0 r_0^{-2}
\delta(\cos \theta - \cos\theta_i) \delta(\phi- \phi_i)$ .
Next we
introduce the function describing the formfactor of impurity in the
form $\delta_a({\Omega}) = (2\pi a)^{-1} e^{(\cos\theta -1)/a}$, the
limit $a\ll 1$ is assumed. The physical range of the potential is
$r_0\sqrt{a}$.
To exponential accuracy we have $\int
d\Omega \delta_a({\Omega}) =1$.  Then, using the asymptotic (Macdonald)
formula for the Legendre polynomials at small $\theta$, we calculate
the auxiliary integral
     \begin{equation}
     \Delta_l \equiv
     A_l^{-1}\int d\Omega Y_{l0}(\Omega) \delta_a({\Omega})
     \simeq
     \exp\left( -a \frac{l(l+1)}{2}\right).
     \end{equation}
This equation shows that as long as considered moments $l<1/\sqrt{a}$, one can
approximate the potential by delta function. 

The impurity potential centered at the north pole reads as
     \begin{equation}
     u(\Omega) = u_ir_0^{-2} \delta_a({\Omega}) =
     u_0r_0^{-2} \sum_l A_l\Delta_l Y_{l0}(\Omega)
     \end{equation}
and the potential centered at the point $\Omega_i$ is
     \begin{equation}
     u(\Omega-\Omega_i) = 
     u_0r_0^{-2} \sum_{lm}
     \Delta_l
     Y_{lm}^\ast(\Omega_i) Y_{lm}(\Omega).
     \end{equation}

\noindent
The averaging over the disorder $\overline{(\ldots)}$
is written as
     \begin{eqnarray}
     \overline{u(\Omega_{1i})
     u(\Omega_{2i})}
     &=&
     \nu_{imp} r_0^2
     \int d\Omega_i u(\Omega_{1i}) u(\Omega_{2i})
     \\
     &=&
     \nu_{imp} u_0^2 r_0^{-2}\sum_l
     \Delta_l^2 A_l
     Y_{l0}(\Omega_{12})
     \\
     &=&
     \nu_{imp} u_0^2 r_0^{-2}\delta_{2a}({\Omega_{12}})
     \end{eqnarray}
Here $\nu_{imp}= N_{imp}/(4\pi r_0^2)$
is the impurity concentration and $N_{imp}$ is the total number of
impurity centers.

Consider next more complicated objects, which are to be used below.
If we have two functions,
$f(\Omega) = \sum_{lm} f_{lm}A_l Y_{lm}(\Omega) $ and
$g(\Omega) = \sum_{lm} g_{lm}A_l Y_{lm}(\Omega) $, then
     \begin{eqnarray}
     f(\Omega) g(\Omega)
     &=&
     \sum_{l_i,m_j}
     (f_{l_1m_1}g_{l_2m_2}
     V^{l_3m_3}_{l_1m_1l_2m_2})
     A_{l_3}Y_{l_3m_3}(\Omega)
     \nonumber     \\
     V^{l_3m_3}_{l_1m_1l_2m_2}
     &=&
     A_{l_1}^2 A_{l_2}^2 A_{l_3}^{-2}
     C^{l_30}_{l_10l_20}
     C^{l_3m_3}_{l_1m_1l_2m_2}
     \end{eqnarray}
with the Clebsch-Gordan coefficients $C^{lm}_{l_1m_1l_2m_2}$.  \cite{VMKh} 

Particularly, when $ f(\Omega) = \delta_a({\Omega})$
and $g(\Omega)$ can be approximated by a constant $\tilde g$ at small
$\Omega \leq \sqrt{a}$, we have $(f \cdot g)_{lm}  = \tilde g
\Delta_l \delta_{m0}$.

\subsection{perturbation theory}

\begin{figure}[ht]
\includegraphics[width=0.8\linewidth] {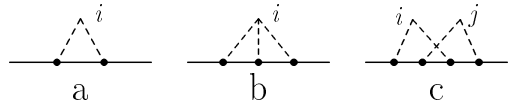}%
\caption{First diagrams describing impurity scattering  }
\label{fig:series} 
\end{figure}

We use the model of point-like impurities randomly distributed on
the sphere. The amplitudes $u_0$ of the impurity potential are chosen to be
equal.   The overall potential is 
     \begin{equation}
     U (\Omega) = 
     u_0r_0^{-2} \sum_{lm}
     \Delta_l Y_{lm}^\ast(\Omega) \sum_{i=1}^{N_{imp}}
     Y_{lm}(\Omega_i) 
     \end{equation}
As usual,  the average of this potential gives a constant, which is
incorporated into the chemical potential  and is
discarded  below.  First nontrivial  diagrams in the perturbation series in
$u_0$ are shown in Fig.\ \ref{fig:series}. 

In the self-consistent Born approximation, we find for the self-energy
in the second order of $U$, Fig.\ \ref{fig:series}a
     \begin{equation}
     \Sigma^{(a)} (\Omega) =
     \nu_{imp} u_0^2 r_0^{-2}
     \delta_{2a}(\Omega) G(\Omega)
     \end{equation}
so that $\Sigma_l^{(a)}$ is independent of $l$ (as long as  $\Delta_l \simeq1$)
     \begin{equation}
     \Sigma_l^{(a)}=
     \nu_{imp} u_0^2
     G(\Omega\sim \sqrt{a})
     \end{equation}
The last line is obtained upon an assumption $L \sqrt{a} <1$, when the
argument of the logarithm in (\ref{G0small}) is small. Recalling that
$L \leftrightarrow k_Fr_0$, the latter condition means that the
range of the impurity potential does not exceed $k_F^{-1}$ in real
space.

Letting $\omega=0$ and introducing the
quantity $$z = \pi \Sigma_l /\Delta E$$
we find the self-consistency equation in this order  of $U$ :
     \begin{eqnarray}
     z  &\simeq&
     X \tilde u ^2
     \left(
     \frac2\pi \ln(L\sqrt{a}) - \cot z
     \right),
     \label{sigma2}
     \end{eqnarray}
with
     \begin{equation}
     X =  N_{imp} /(2L+1),
     \label{extent}
     \end{equation}
and the dimensionless scattering amplitude
     \begin{equation}
     \tilde u = u_0 m_e/2.
     \label{strength}
     \end{equation}
In the following we adopt the picture of the weak scattering,
$\tilde u \ll 1$.

The real part of $\Sigma_l$ in (\ref{sigma2}) redefines the
chemical potential and may be ignored.  The imaginary part of
$\Sigma_l$ describes the width of the impurity-induced band, arising instead
of degenerate  multiplet $L$ : 
     \begin{eqnarray}
     z &=&
     X \tilde u ^2
     \left(
     \frac2\pi \ln(L\sqrt{a}) - i
     \right)
      , \quad X \tilde u ^2\agt 1
     \label{largeX}
     \\ &=&
     -i \sqrt{X} \tilde u , \quad X \tilde u ^2\alt 1
     \label{smallX}
     \end{eqnarray}
As we discussed above,  the second regime is of our main interest. It
corresponds to the case when the $\cot z$ dominates over the logarithm in
(\ref{sigma2}), which in turn means the possibility to ignore the transitions
between the levels with different $l$. The scattering in this case is between
the states with different $m$ within one multiplet $l$.

The next  diagram,  Fig.\ \ref{fig:series}b,  is wrtten in the form
     \begin{equation}
     \Sigma^{(b)} (\Omega) \simeq
     \nu_{imp} u_0^3 r_0^{-2}
     \delta_{a}(\Omega)
     G^2(\Omega).
     \end{equation}
Or, in the previous notation
     \begin{equation}
     \frac{\pi\Sigma^{(b)}_l}{\Delta E }
     \simeq
     X \tilde u ^3
     \left(
     \frac2\pi \ln(L\sqrt{a}) - \cot z
     \right)^2.
     \label{sigma4b}
     \end{equation}

In order to find the relative importance of this correction,  we
put the former estimate, $\Sigma^{a}$, into the energy argument of the
Green function, $\lambda \simeq l+1/2-\Sigma_l^{a}/\Delta E $.
Then we find
     \begin{eqnarray}
     {\Sigma^{b}_l}/{\Sigma^{a}_l}  &\sim&
     \tilde u 
     , \quad X \tilde u ^2\agt 1,
     \label{sigmaBlarge}
     \\ &\sim&
     X^{-1/2} , \quad X \tilde u ^2\alt 1.
     \end{eqnarray}
Evidently, the correction $\Sigma^{b}$ can be neglected at large $X$
and  dominates at $X<1$. We discuss this feature in more detail below. 

The next diagram is shown in the Fig.\  \ref{fig:series}c
     \begin{equation}
     \Sigma^{c} (\Omega) =
     \nu_{imp}^2 u_0^4
     G(\Omega)^3
     \end{equation}
The Fourier component $\Sigma^{c}_l$ behaves differently in the two
above cases $z>1$ and $z<1$.

In the case $z\ll 1$ the Green function can be approximated by the
expression (\ref{G0series}). Near the level $E_L$ we have
     \begin{eqnarray}
     G &\simeq&
     (4\pi r_0^2)^{-1} \frac{ (2L+1) P_L(\cos\Omega)
     }{\omega +\mu -E_{L}-\Sigma_L} ,
     \label{G-Legendre}
     \\   &\simeq&
     - \frac{m_e} {z}
     \frac{\cos[(L+1/2) \Omega-\pi/4]}
     {\sqrt{2\pi L\sin\Omega}} , \quad L\sin\Omega \agt 1
     \label{G-Leg-asy}
     \end{eqnarray}
where we have put $\omega=0$ in the last line. Using
(\ref{G-Legendre}) and (\ref{G-Leg-asy}), one finds for $\Sigma^{c}_l$
the following expressions, respectively, 
     \begin{eqnarray}
     \Sigma^{c}_l &\simeq&
     -\nu_{imp}^2 u_0^4 r_0^2
     \left(\frac{\pi m_e}{z Lr}\right)^3
     \sum_{l_1}
     V^{L0}_{l_10l0}
     V^{l_10}_{l0l0},
     \label{sigma4c1}
     \\
     &\simeq&
     - \frac{3}{2}\nu_{imp}^2 u_0^4
      A_l^{-1}
     \left(\frac{m_e}{2\pi zA_L }\right)^3
     \ln\left|
     \frac{L}{L-l}
     \right|.   ,
     \\
     &\simeq&
     - \Delta E \frac{3 X^2 \tilde u^4}{\pi^3 \sqrt{lL} z^3}
     \ln\left|
     \frac{L}{L-l}
     \right|.       .
     \label{sigma4c2}
     \end{eqnarray}
Note that the expression (\ref{sigma4c1}) tells us that : i)
$l$ and $L$ are even (odd) simultaneously, i.e. $(L+l)=0{\rm mod} 2$,
and ii) $\Sigma^{c}_l = 0$ at $l>3L$. The Eq.\ (\ref{sigma4c1}) is not
convenient and one can use an approximate expression (\ref{sigma4c2}).  
%
%
In the considered  case (\ref{smallX}) we estimate at $l=L$ :
     \begin{eqnarray}
     {\Sigma^{c}_L}/{\Sigma^{a}_L}  &\sim&
     L^{-1} \ln L \ll1
     \end{eqnarray}
Therefore this correction is smaller than the previous one. 

The case $Im\, z>1$ corresponds to the mean free path of the electron
smaller than $r_0$. In this case the Green function (\ref{G0-asymp}) 
exhibits the presence of only one wave, with the shorter path. One has from
(\ref{G0-asymp}) at $Im \lambda \agt 1$ :
     \begin{equation}
     G \simeq
     - \frac{m_e}{\sqrt{2\pi \lambda \Omega}}
     \exp\left[i \lambda \Omega+i\frac\pi4 \right],
     \label{G0-damped}
     \end{equation}
As a result we have
     \begin{eqnarray}
     \Sigma^{c}_l &\simeq&
     \Delta E \frac{X^2 \tilde u^4}{ L}
     f\left( \frac{l+1/2}{3 \lambda} \right)
     \label{sigma4c3}
     \\
     f(x) &=& \frac{(2/\pi)^3 }{\sqrt{6(1+x)}}
     K\left[\frac{2x}{1+x} \right].
     \end{eqnarray}
with the complete elliptic integral $K(x)$ and $\lambda \simeq L+1/2 +
z/\pi$. A comparative importance of this contribution is
     \begin{eqnarray}
     {\Sigma^{c}_L}/{\Sigma^{a}_L}  &\sim&
     X \tilde u^2 /L
     \label{sigmaClarge}
     \end{eqnarray}
and may be significant in the considered case $X \tilde u^2 >1$. 

\begin{figure}[ht]
\begin{center}
\includegraphics[width=0.4\linewidth] {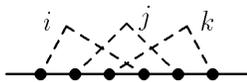}%
\caption{A sixth-order diagram with maximally crossed impurity lines}
\label{fig:6th} 
\end{center}
\end{figure}

One can evaluate also a  sixth-order correction,  shown in the Fig.\ 
\ref{fig:6th},  which is the next non-trivial term.    In case
of our interest, $X\tilde u^2 \alt 1$, the contribution of this diagram,
$\Sigma^{d}$, is estimated as follows.
     \begin{eqnarray}
     \Sigma^{d}_{l=L} &\simeq&
     \Delta E
     \frac{X^3 \tilde u^6}{z^5}
     \frac{4}{\pi^3 L}
     \end{eqnarray}
so that the relative importance of this diagram is again
small :      
     \begin{equation}
     {\Sigma^{d}_L}/{\Sigma^{a}_L}  \sim 1/L
     \end{equation}

One can further argue that the weak localization corrections do not
contribute much at $X\tilde u^2 <1$. Indeed, in this
case of our interest below, when the mean free path is larger than the radius,
we are essentially in the ballistic regime, hence one does not find a typical
enhancement in the cooperon series. As a result, the arguments elucidating the
role of the spherical topology in the diffusive regime \cite{Kravtsov99}
cannot be fully applied to our case. 

On the other hand, let us consider a diagram which describes four scattering
events on the same impurity.  This is the next diagram in a  sequence, formed
by  Figs.\ \ref{fig:series}a,  \ref{fig:series}b.  It can be easily shown,
that in the regime (\ref{smallX}) this diagram is estimated as
$\Sigma^{a}_L/X$, being thus large at $X<1$.  This latter sequence of diagrams
contributes to the S-matrix for one impurity.  The formal divergence in 
this sequence  at $X<1$  indicates a  singularity in the S-matrix. In the
following section we show, that this singularity corresponds to the 
 incomplete raising of degeneracy in $L-$th multiplet. 

Note that the quantities $L$, $X$ and $(\Delta E)^{-1}$ scale as $r_0$
so that in the limit of large sphere the regime (\ref{smallX}) does not occur
and we are left with eqs.\ (\ref{largeX}), (\ref{sigmaBlarge}),
(\ref{sigmaClarge}).
It is worth to represent the quantity $X$ in the form
        \begin{equation}
        X = \frac{\nu_{imp}}{\sqrt{\nu/\pi}}r_0,
        \end{equation}
then we see that the criterion $X<1$  means  $r_0 \alt
\sqrt{\nu}/\nu_{imp}$. For the semiconducting situation, adopting a simple
estimate $\nu\sim\nu_{imp}\sim 10^{10}\,$cm$^{-2}$, we get for this
regime $r_0 \alt 100\,$nm, the latter value comparable to the size of
opal balls.

Concluding this section, we observe that at sufficiently small disorder,
$X\tilde u^2 <1$, the impurity-induced bands, referring to different
multiplets, do not overlap. This regime corresponds to the scattering mostly
within one multiplet. At small $X$ one meets a singularity in S-matrix of
scattering for one impurity.  We  consider it to some detail in the
next section.

\section{ scattering within a multiplet}

In this section we restrict ourselves by the analysis of the 
scattering within a multiplet, characterized by the quantum number $L$.
Initially this multiplet is $(2L+1)-$fold degenerate. The 
impurities lower the rotational symmetry and lift the degeneracy.
The transitions between the multiplets with different $L$ are
neglected.  The matrix elements of the Hamiltonian are the projection
of the impurity $\delta-$functions onto the multiplet, and we have
     \begin{eqnarray}
     {\cal H}_{m,n} 
     &=&
     \sum_{i=1}^{N_{imp}}
     u_i r_0^{-2}
     Y_{Lm}^\ast(\Omega_i)
     Y_{Ln}(\Omega_i)
     \label{Ham-L}
     \end{eqnarray}

In this section, in order to obtain a wider picture,  we allow a variation
of $u_i$.  Specifically, we numerically consider  three possibilities ~: i) the
former case,  all $u_i=u_0 $, ii) $u_i=\pm u_0$ and $\overline{ u_i }= 0 $,
and iii) $u_i$ is a normally distributed (ND) variable with variation $u_0$,
$\overline{ \exp(ik u_j)} = \exp (-k^2u_0^2/2)$.

Simple estimates show that the average off-diagonal matrix element of
$\cal H$ scales in the limit of large $N_{imp}$ as
     \begin{equation}
     \overline{|{\cal H}_{m,n}|^2} \sim
     u_0^2 r_0^{-4} N_{imp}
     \end{equation}
In this limit one may also regard different ${\cal H}_{m,n}$
to be almost statistically independent quantities. Due to the known
property of large random matrices, it follows then that the width of
the impurity-induced band will be of order of $u_0 r_0^{-2}\sqrt{N_{imp}
L}$. This estimate corresponds to the above one, eq.\ (\ref{smallX}),
and is validated by the numerical calculations. At the same time,
some features  of our problem prevent us from an identification of ${\cal
H}$ with one of the classes of random matrices.  Hence  the usual 
expectations from the random matrix theory might not be fully applicable to our
case. We discuss these points in the next subsections. 

\subsection{Lifshits's theorem}

There exists a particular property, satisfied when
the degeneracy $(2L+1)$ is larger than the number of impurity centers,
$N_{imp}$. In this case at least $2L+1-N_{imp}$
levels remain degenerate for any choice of $u_i, \Omega_i$.
This statement is a general one for the projective perturbations
and dates back to the work by Lifshits \cite{Lifshits47}. It is
known particularly for the impurity scattering of the electrons in a
magnetic field.\cite{Ando,baskin}  For a sphere, this statement can be proved 
as follows.

Let us write $Y_{Lm}(\Omega_i)
\equiv \langle i| m\rangle$ and introduce a scalar product for two
points on the sphere as
        \begin{eqnarray}
        \langle i| j\rangle &=&
        \sum_m
        Y_{Lm}^\ast(\Omega_j)
        Y_{Lm}(\Omega_i) =
        \frac{2L+1}{4\pi} P_L(\cos\Omega_{ij})
        \nonumber
        \end{eqnarray}
We introduce also a real-valued matrix ${\cal M}$ with elements
        \begin{equation}
        {\cal M}_{ij} = u_i r_0^{-2} \langle i| j\rangle.
        \label{Ham-M}
        \end{equation}
Then
${\rm Tr} {\cal H} = \sum_{im}
u_i r_0^{-2} \langle m| i\rangle \langle i| m\rangle = {\rm Tr} {\cal
M}$, and
${\rm Tr} {\cal H}^2 = \sum_{ijmn}
u_i  u_j r_0^{-4}
\langle n| i\rangle \langle i| m\rangle
\langle m| j\rangle \langle j| n\rangle
= {\rm Tr} {\cal M}^2$, and generally
        \begin{equation}
        {\rm Tr} {\cal H}^k = {\rm Tr} {\cal M}^k.
        \end{equation}
After this observation it is straightforward to show that
        \begin{eqnarray}
        det(E-{\cal H})      &= &
        \exp\left[ (2L+1) \ln E +
        {\rm Tr} \ln(1 -{\cal H}/E)
        \right]
        \nonumber \\  &= &
        \exp\left[ (2L+1) \ln E +
        {\rm Tr} \ln(1 -{\cal M}/E)
        \right]
        \nonumber \\  &= &
        E^{2L+1- N_{imp}}
        det(E-{\cal M}).
        \end{eqnarray}
In the last line the pole in $E$ at $2L+1 < N_{imp}$ is compensated by
the necessary degeneracy of the matrix $\cal M$ in this case.

Therefore a singularity in the S-matrix of perturbation theory is identified
with the incomplete raising of degeneracy. 

Another specific feature of our problem is the following.  We see that
non-zero eigenvalues of  (\ref{Ham-L}) coincide with those of
(\ref{Ham-M}) which is a real symmetric matrix (for $u_i=u_0$) with a
randomness in its elements.   Then one should seemingly classify $\cal M$ to
Gaussian orthogonal ensemble (GOE).  However, the elements of $\cal M$ are not
statistically independent. Indeed,  it can be easily shown, that averaging 
over positions of impurities gives 
          \[
            \overline{{\cal M}_{ij}^2}  = 
            \frac{(2L+1) u_0^2}{(4\pi)^2 r_0^4}\equiv Q^2 ,
          \]
while, e.g., the triple combination with the cyclic sequence of indices yields
          \[
          \overline{{\cal M}_{ij} {\cal M}_{jk} {\cal M}_{ki} } =
          \frac{Q^3}{\sqrt{2L+1}}      \neq 0.
         \]
This fact of nonvanishing long-range correlations makes our problem distinct
from the usually considered ones.  Being small, these correlations however
should cause only minor deviations  from the GOE predictions. 

For the case, when the amplitudes $u_i$ are sign-reversal, $\cal M $ is not
symmetric and  one expects larger deviations from the known picture. 

\subsection{numerical results}

We performed the numerical diagonalization of the matrices of the form
(\ref{Ham-L}),
with the positions $\Omega_i$  randomly distributed on the
sphere and the amplitudes $u_i$  chosen according to one of the above
ensembles. The calculations were done with the use of standard EISPACK
routines for $10,000$ random realizations of $\Omega_i,u_i$ for each
$L,N_{imp}$.

\begin{figure}[ht]
\begin{center}
\leavevmode
\includegraphics[width=0.8\linewidth] {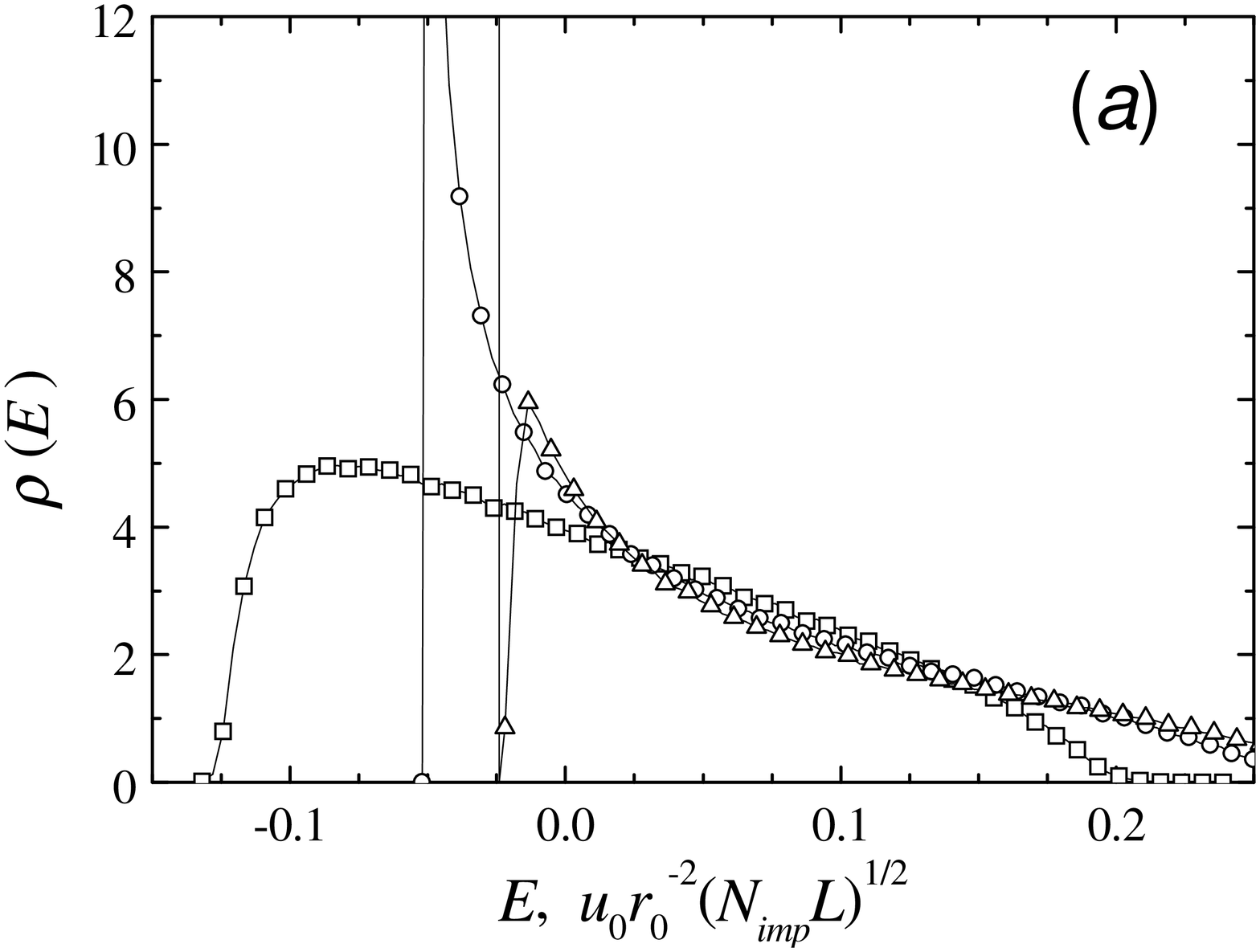}\vskip0.8cm
\includegraphics[width=0.8\linewidth] {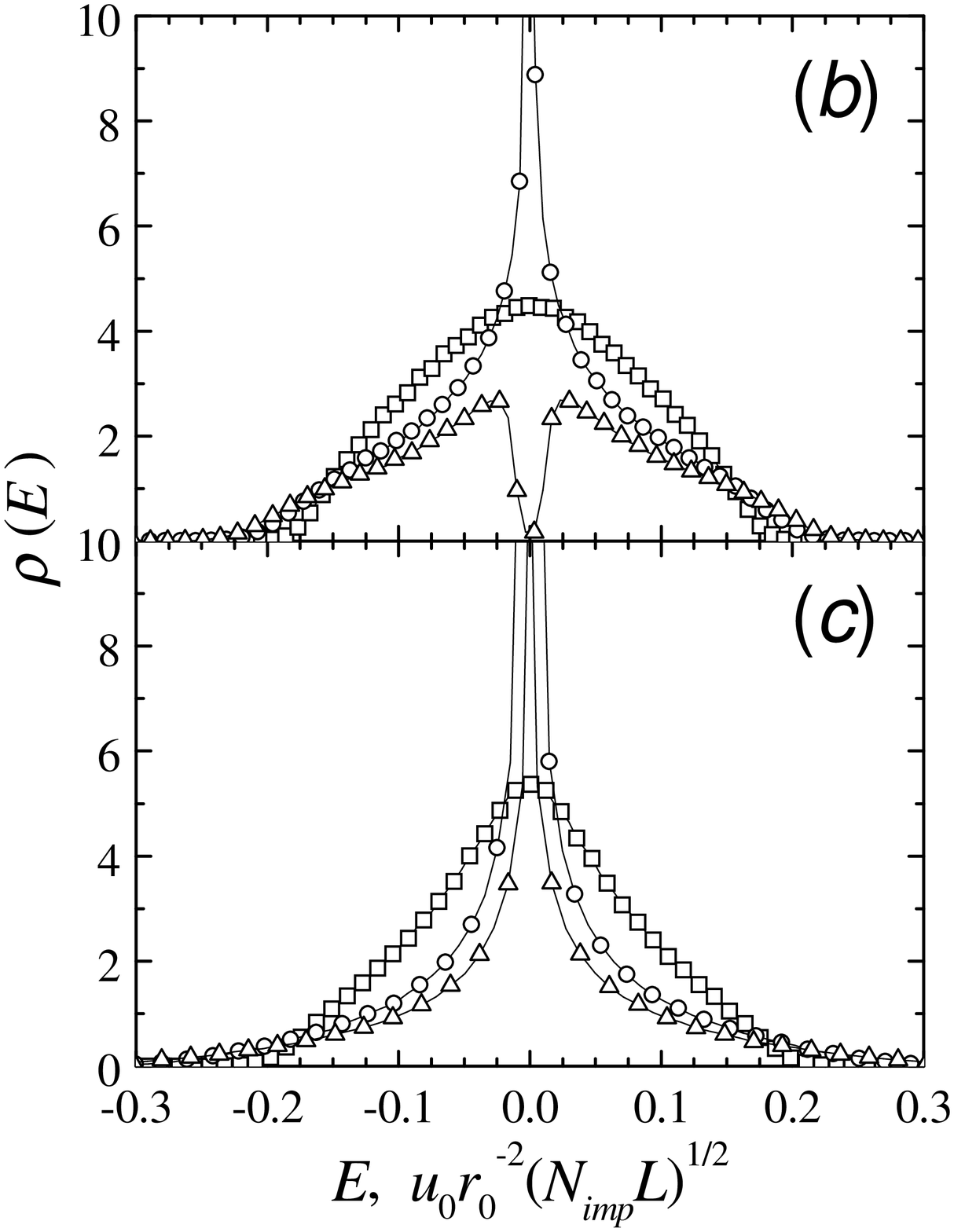}%
\caption{The average density of states as a function of energy,
measured in units discussed in the text.
The data for identical (a), sign-reversal (b) and ND amplitudes (c)
are shown. The values of $X$ are $X=4.92\, ( \Box)$, $1\, (\circ)$, and
$0.62\, (\triangle)$, For convenience of presentation of data with
different $L$, the shown DOS is scaled as $\int \rho(E)\, dE=1$ here.  The
lines are guides to the eye, every second data point is shown.  The
vertical line on the plot (a) denotes the position of the
$\delta-$function for $X=0.62$.}
 \label{fig:dos-u} 
\end{center}
\end{figure}

For given $L,N_{imp}$, we determine the density of states $\rho(E)$ (DOS),
averaging over the realizations.
The density is normalized as $\int_{-\infty}^\infty \rho(E)\,dE =2L+1$ .
Some typical curves for DOS are shown in the Fig.\ \ref{fig:dos-u}. 
We observe following qualitative features : 

\noindent  i) For identical  impurities, $u_i=u_0$,  we have a smooth
asymmetric  DOS at $X>1$, when the degeneracy is removed.  At the point $X=1$, 
DOS exhibits a mild singularity, roughly of the law $\ln^2 E$. When $X<1$, a
part of the multiplet is unsplit, and the split-off states are separated from
the position of this $\delta-$function.  For a convenience of
presentation of Fig. \ref{fig:dos-u}a, we shifted the energies by
$\mbox{Tr} {\cal H}/(2L+1) = \nu_{imp} u_0$. As a result,  the
DOS is centered for each $X$, i.e. $\int_{-\infty}^\infty E\, \rho(E)\,dE =0$. 

\noindent ii) For the ``dichotomic'' impurities, $u_i=\pm u_0$, $\overline{
u_i} =0$, the DOS is symmetric, showing the same features as in the case (i). 

\noindent iii) For the case of the normally distributed $u_i$, the properties
of DOS are similar to the case  (ii), except for the region $X<1$. In this
latter region,  $\rho(E)$ shows a mild singularity at $E=0$,
accompanied by  $\delta-$function contribution of unsplit states. 

In order to test the accuracy of calculations, we fixed the window for the
expected $N_0 = 2L+1-N_{imp}$ unsplit states to the size $10^{-6}$ of estimated
band-width. 
The average number of energy levels, found in this window was  $N_0+\delta N$
with $\delta N\ll 1$.  In the worst case of ND amplitudes $u_i$ and $X<1$ we
had $\delta N\alt 10^{-3}$.  The small values of $\delta N$ allowed us to
separate the smooth contribution from the $\delta-$core. 

As a next step, we unfolded  the spectrum, $\{E_i\}\to \{\ve_i\}$ by defining
$\ve_i = \int_{-\infty}^{E_i} \rho(E)\,dE$. A certain subtlety in this
procedure should be described here.  If the degeneracy of the multiplet is
completely removed, then the average spacing $s_i \equiv \ve_{i+1}-\ve_i$ 
between the nearest neighboring $\ve_i$ is $\overline{s_i}= 1$.
We examine also the statistics of the spectrum for the partially
degenerate level when the DOS contains a smooth part and a
singular contribution, $\rho(E) = \rho_{smooth}(E) + N_0 \delta(E)$. To
describe this case, we exclude the states in the  $\delta-$core from our
consideration  and write for the split levels  $\ve_i = \int_{-\infty}^{E_i}
\rho_{smooth}(E)\,dE$. Obviously, upon doing this, we have again  $\overline{(
\ve_{i+1}-\ve_i)}=1$, now with $i=1,\ldots ,N_{imp}-1$. 

We studied the energy level spacing distribution function $P(s)$ with the use
of the unfolded spectrum. 
We remind, that usually in analyzing the spectrum of random ensembles, one
finds either the Wigner-Dyson (WD) law, $P(s)\sim s \exp (-s^2)$, established
for the GOE ensemble, or the Poissonian law, $P(s) \sim
e^{-s}$.  
Besides, it was shown that in the vicinity of the 
metal-insulator Anderson transition (MIT) an intermediate distribution may take
place, which interpolates between the WD and Poissonian dependence.
\cite{Aronov95}
Particularly, one observes a 
``semi-Poisson'' distribution,  $P(s) = 4s e^{-2s}$ in three dimensions (D) at
MIT. \cite{boundary}
Similar intermediate $P(s)$ occur also in the studies of the statistics of
the lowest Landau level in the presence of disorder \cite{Feingold} and in
some other models. \cite{Berkovits}

\begin{figure}[ht]
\includegraphics[width=0.8\linewidth] {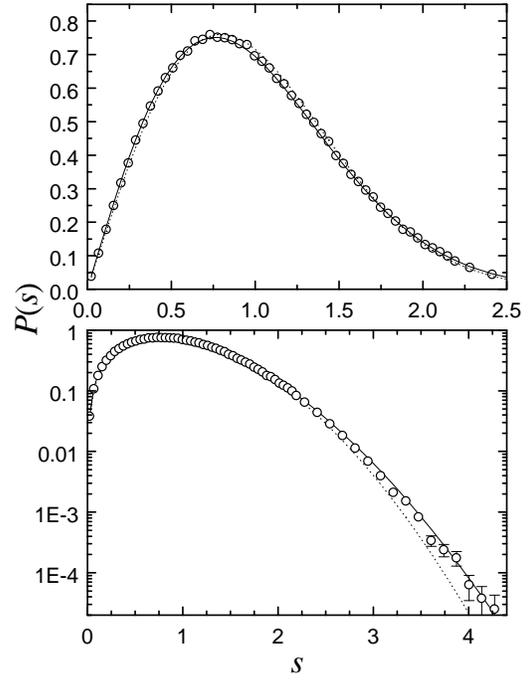}%
\caption{
Energy spacings distribution for $L=61$, $N_{imp}=300$, identical
impurities; $b=1.88$ .}
 \label{fig:p61300}  
\end{figure}

\begin{figure}[ht]
\includegraphics[width=0.8\linewidth]{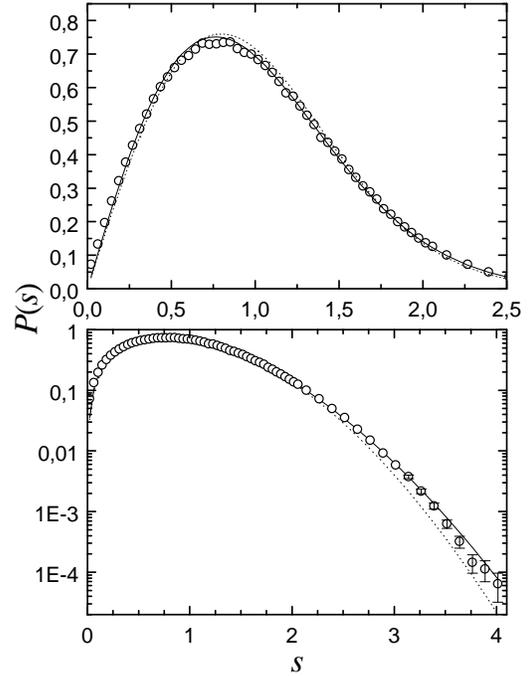}%
\caption{
Energy spacings distribution for $L=40$, $N_{imp}=50$, identical
impurities; $b=1.82$ .}
\label{fig:p81050}  
\end{figure}

\begin{figure}[ht]
\includegraphics[width=0.8\linewidth]{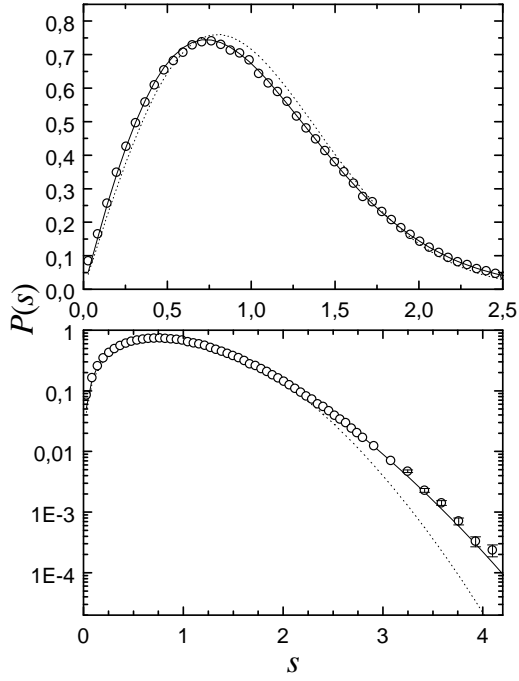}%
\caption{
Energy spacings distribution for $L=30$, $N_{imp}=50$, ``dichotomic''
impurities; $b=1.68$ .}
 \label{fig:p6105m}  
\end{figure}

\begin{figure}[ht]
\includegraphics[width=0.8\linewidth] {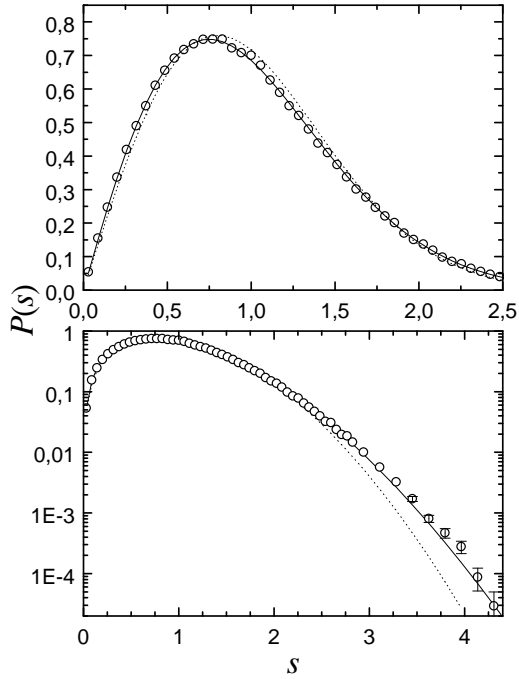}%
\caption{
Energy spacings distribution for $L=20$, $N_{imp}=200$,  ND
impurities; $b=1.76$ . } \label{fig:p4120g} 
\end{figure}

It turned out, that our data are well described (with  some predictable
exclusions, discussed below) by the following formula
           \begin{equation}
           P(s) = cs\, \exp(-a s^b).
           \label{modWD}
           \end{equation}
Since the conditions $\int P(s)ds = \int s\,P(s)ds =1$
uniquely determine the parameters $a,c$ in the form
        \begin{eqnarray}
        a &=& \left({\Gamma[3/b]}/{\Gamma[2/b]} \right)^b,
        \\
        c&=& b\, \Gamma[3/b]^2 \,\Gamma[2/b]^{-3} \nonumber,
        \end{eqnarray}
the eq.\ (\ref{modWD}) depends, in fact, only on $b$. This parameter was
determined by fitting the numerically obtained curves both for $P(s)$ and $\ln
P(s)$.  Remarkably, in those cases, when a good visual agreement of the data
with a fitting curve  was achieved, the values of $b$, determined both ways,
essentially coincided.   Some of the results are shown in Fig.\ 
\ref{fig:p61300},  \ref{fig:p81050},  \ref{fig:p6105m},  \ref{fig:p4120g}.  

The values of $b$ are  summarized  in the Table  \ref{table1}.  In cases of
good visual agreement,  the corresponding error  in $b$ was $\pm 0.01$, 
estimated from the different fitting procedures used. 
From this table, we see that the values of $b$
in the investigated range of $L$, $N_{imp}$ depend only on the ratio of these
quantities.  The obtained $b$s are essentially the same in each of two domains
$X>1$ and $X<1$, with  somewhat lower $b$s  at $X<1$. The worse agreement with
the fit was obtained in the cases, when the regular part of DOS,
$\rho_{smooth}(E)$ showed a mild singularity, Fig.\  \ref{fig:dos-u}.   In
these cases we checked  the procedure of  unfolding the spectrum, which
included the cubic spline interpolation. The range of $E$ was divided, so that
the position of singularity corresponded to a free boundary of a spline. 
 The resulting spline curve fitted $\rho_{smooth}$ well,
showing no oscillations , however, it did not improve the quality of the fit
of $P(s)$ by eq. (\ref{modWD}).  Particularly, the tail of $P(s)$ in these
cases showed roughly exponential decay $\ln P(s) \sim  - s $ at larger $s$ and
a small non-vanishing value of $P(s=0)$,  which can be expected for the 
systems with a localization of the  states. The results of the  fit of this
tail  by  the  exponential law for the normally distributed  $u_i$ are shown
in Fig.\   \ref{table2}.  It is seen here, that $P(s)$ at large $s$  
decreases faster with the increase of disorder $X$. 

The linear tail in $\ln P(s)$ at large $s$ for the singular DOS can be,  in
principle, expected.  One can argue that in this case the unfolding procedure
itself is not particularly useful. Indeed, this procedure assumes that
different parts of the spectrum can be treated on equal grounds. This
assertion can be tolerated for unsingular DOS, while the presence of a
singularity in $\rho_{smooth}$ explicitly breaks the premise for the
unfolding, $E_i \to \ve_i$. As a result, one anticipates that the part of the
spectrum at the center of the band shows the intermediate statistics, which is
different from the Poissonian one, stemming from the localized states at the
tails of the band (see, e.g., Ref.\ \cite{Feingold}). 

Note, that the considered cases of dichotomic and ND amplitudes lead to the
same value of $b=1.76 \pm 0.02$ at $X\agt 3$. By analogy with the integer
quantum Hall effect (IQHE) \cite{Huckestein} one can argue that the limit
of large $X$ corresponds to  the white noise distribution of impurity
potential ( $X$ is associated with the number of impurities per one flux
quantum).  Our findings suggest that $b\neq 2$ in this limit. 

Next, we performed the analysis of the  number variance, 
             \begin{equation}
             \Sigma^2(\ve) = \overline{N^2(\ve)} 
             - \left( \overline{N(\ve)}\right)^2 ,
             \end{equation}
which measures the fluctuation of the number of levels $N(\ve)$ in a band of
unfolded spectrum of width  $\ve$.  The number variance  is generally believed
to be more sensitive than $P(s)$ to the change of ensemble
statistics. One has $\Sigma^2(\ve)= \ve$ for the Poisson sequence, while GOE
prediction is  $\Sigma^2(\ve)\sim 2 \pi^{-2} \ln \ve$ at large $\ve$.  
For the
intermediate statistics at MIT, earlier theoretical arguments
\cite{Kravtsov94,Aronov94} suggested that $\Sigma^2(\ve) \propto \ve^{2-b}$, 
later it was agreed that  $\Sigma^2(\ve) \sim \chi \ve$ with $\chi <1$.
\cite{AronovMirlin,Kravtsov95,Chalker96}

\begin{figure}[ht]
\includegraphics[width=0.8\linewidth] {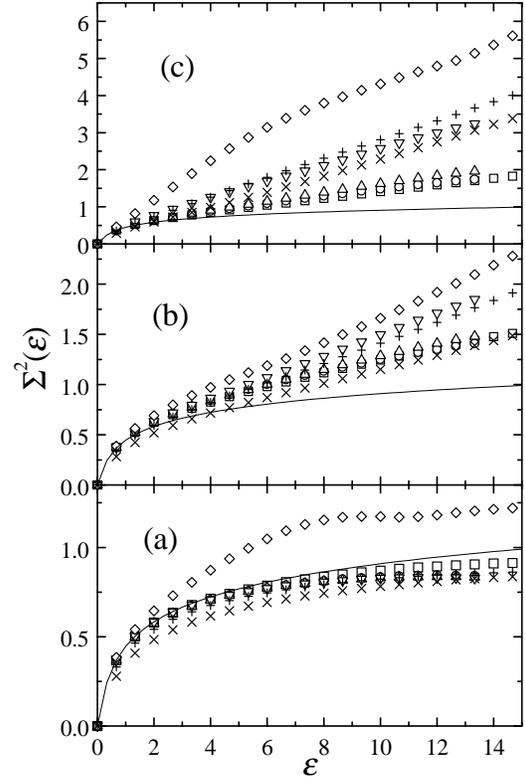}%
\caption{
Number variance for the amplitudes with (a) fixed value, (b) ``dichotomic''
and (c) normal distributions. The values of $X$ are :  $4.92\,
(\Box)$, $4.88\, (\circ)$, $2.93\, (\triangle)$, $1.22\,
(\bigtriangledown)$, $1.0\, (\Diamond)$, $0.82\, (+)$, $0.62\,
(\times)$. The GOE prediction for the number variance is shown by a
solid line.}
\label{fig:sig-all}  
\end{figure}

Since we deal with finite size matrices, the range of $\ve$ is restricted
from above by $(2L+1)$. Actually the finite-size effects are felt
at $\ve \sim 0.4 (2L+1)$, where one finds a  maximum of
$\Sigma^2$.  The behavior  of the number variance at  lower $\ve$ is shown 
in Fig. \ref{fig:sig-all}.  

From this figure one sees that in the case (i) of identical impurities
$\Sigma^2$ follows the GOE prediction except for the value  $X=1$. 
In cases (ii) and (iii), when $u_i$ are allowed to vary, the dependence
$\Sigma^2(\ve)$ goes visibly above the GOE law.  One notices that in the case
(iii) of ND $u_i$ the obtained  $\Sigma^2$ is considerably higher than in the
case (ii) of dichotomic amplitudes.  The highest curve for the number
variance is obtained again at $X=1$, both for the case (ii) and (iii). 

The pronounced variation of $\Sigma^2$ with $X$ in two latter cases should be
contrasted with our results for $P(s)$, Table \ref{table1}.  Indeed, we see
that the the curves for $\Sigma^2$ may almost coincide for different $b$'s
(at $X<1$ and $X>1$), while  they may be notably different for almost
identical $b$'s.  Particularly, one  finds almost linear dependence at $\ve  \agt 3$ , 
$\Sigma^2(\ve) \simeq \chi \ve +const$ with $\chi <1$.  The slopes $\chi$ vary
in the range $0.06 \div 0.14$ and $0.1 \div 0.4$ for the dichotomic and ND
amplitudes, respectively. 

Making this observation we try to compare our system with the  problems  of
metal-insulator Anderson transition and of impurity-split lowest Landau level.
We mentioned the earlier theoretical expectation $\Sigma^2(\ve) \propto
\ve^{2-b}$ above. This possibility, numerically validated for the 3D MIT
\cite{Evangelou} is apparently not favored by our data. 
Other studies support the viewpoint that $\Sigma^2(\ve)
\propto \chi N$ with the level compressibility $\chi<1$  (see
\cite{Mirlin-hab} for a review).    A relation $ 2\chi = 1 -D_2/d $ connects
\cite{Chalker96,MirlinEvers} the quantity $\chi$  with the multifractal
exponent $D_2$, characterizing the spatial extent of the wave functions in the
flat geometry of $d$ dimensions, $\int d{\bf r}\,  |\psi|^4 \propto V^{-D_2/d}$
with  $V$  a volume of the system.  
 For the IQHE various authors provided the values
$D_2=1.4 \div 1.6$ and $\chi = 0.1 \div 0.15$.
\cite{Huckestein94,Matsuoka97,Klesse97} 

The values of $\chi$ extracted from our data are roughly consistent with these
results. A pronounced dependence of $\chi$ on $X$, however, requires further
understanding. We  suggest here that \\
 i) either our results obtained  for $\ve
\leq 15$ do not describe the true asymptote of $\Sigma^2$, (cf.
\cite{Evangelou,boundary})  \\
ii) or the multifractal exponent $D_2$ varies
with $X$. 

It would be interesting to check the last possibility by investigating the
spatial character of the wave functions in our model. This work is however
beyond the scope of the present study. 

\section{Discussion and conclusions}

We considered the electron gas on the sphere, moving in the potential field of
point-like scatterers. Without a disorder,  one observes  a
series of effects, associated with the coherent motion of the electrons in
the spherical geometry. The high symmetry of a problem leads to the large
degeneracy of energy spectrum in the ideal case. The disorder removes the
rotational symmetry, prompting one to expect the disappearance of the 
coherence effects.

Our consideration shows  that there exist a definite range of disorder
parameters, where the sphere remains sphere at least partly.
Particularly we demonstrate that the phenomenon of incomplete raising of
degeneracy may take place for sufficiently small radii. In the case of
relatively small disorder the main effect is the splitting of the rotational
multiplet, while the transitions between the different multiplets can be
ignored.  Numerically investigating this latter case, we studied the energy
level statistics in the impurity-induced band.  We found that the energy
spacing distribution function follows a modified Wigner-Dyson law,
$P(s)\sim s \exp(- s^b)$ with $b\simeq 1.76$ in the white noise limit.  Our
data suggest that the parameter $b$  does not determine the behavior of the
energy levels' number variance $\Sigma^2$. The latter quantity reveals 
various  possibilities for almost the same values of $b$, depending strongly
on the chosen model of disorder. 

\begin{acknowledgments}
I thank
D.Braun,  P.A. Braun, K. Busch, K.J. Eriksen,
Y.V. Fyodorov, I.V. Gornyi, A.D. Mirlin, M.L. Titov, 
for various helpful discussions.
The partial financial support from the RFBR Grant 00-02-16873,
Russian State Program for Statistical Physics (Grant VIII-2) and grant
INTAS 97-1342 is gratefully acknowledged.
\end{acknowledgments}

%
%
%


\begin{table}
\caption{
The values of the best fit to $P(s)$ by the formula
$P(s) = cs\, \exp(-a s^b)$. Parameters $a,c$ depend on $b$ as described
in text.
\label{table1}}
\begin{tabular}{cccccc}
$(2L+1)$ & $N_{imp}$ &$X$ & $b (u_i=1)$ & $b (u_i=\pm1)$ & $b (u_i=ND)$
\\  \tableline
61 & 300 & 4.92 & 1.88 & 1.74 &  1.78\\
41 & 200 & 4.88 & 1.89 & 1.76 &  1.76\\
41 & 120 & 2.93 & 1.89 & 1.75 & 1.75 \\
41 & 50  & 1.22 & 1.86 & 1.73 & 1.62 \tablenotemark[1]\\
61 & 61 &1 & 1.64\footnotemark[1] & 1.68  &  1.34\tablenotemark[1] \\
61 & 50 &0.82 & 1.80 & 1.68 &  1.45\tablenotemark[1] \\
81 & 50 &0.62 & 1.82 & 1.68 &  1.43\tablenotemark[1] \\
\end{tabular}
\tablenotetext[1]{ worse visual agreement with $P(s)$,
linear tail of $\ln P(s)$ at $s\agt 2$ }
\end{table}

\begin{table}
\caption{
The values of the best fit to $\ln P(s)$ by the formula
$\ln P(s) = -a_1 -a_2 (s-2)$ at $2<s<5$.  ND amplitudes.
\label{table2}}
\begin{tabular}{ccccc}
$(2L+1)$ & $N_{imp}$ &$X$ & $a_1$ & $a_2$ \\
\tableline
41 & 50 &1.22 & 1.75(6) & 2.75(4)  \\
61 & 61 &1    & 1.81(2) & 2.20(1)  \\
61 & 50 &0.82 & 1.97(5) & 2.12(3)  \\
81 & 50 &0.62 & 2.15(8) & 1.82(3)  \\
\end{tabular}
\end{table}


\newpage

\vskip.8cm

\vskip.8cm


\vskip.8cm

\vskip.8cm

\vskip.8cm

\end{multicols}

\end{document}